\documentclass[a4paper,11pt]{article}

\usepackage{amsmath}
\usepackage{amsfonts}
\usepackage{amssymb}
\usepackage{graphicx}

\newcommand{\ME}[3]{\left\langle #1 \right| #2 \left| #3 \right\rangle}

\newcommand{\skp}[2]{\left\langle #1 \right| \left. #2 \right\rangle}
\newcommand{\st}[1]{\left| #1 \right\rangle}

\newcommand{\bra}[1]{\left\langle #1 \right|}
\newcommand{\op}[2]{\left| #1 \right\rangle \left\langle #2 \right|}

\newcommand{\eqr}[1]{Eq.~(\ref{#1})}
\newcommand{\fir}[1]{Fig.~\ref{#1}}

\def\be{\begin{equation}}
\def\ee{\end{equation}}
\def\bea{\begin{eqnarray}}
\def\eea{\end{eqnarray}}
\def\bes{\begin{equation*}}
\def\ees{\end{equation*}}
\def\beas{\begin{eqnarray*}}
\def\eeas{\end{eqnarray*}}

\begin{document}

\title{Optical Lattices, Ultracold Atoms and Quantum Information Processing}

\author{D.~Jaksch}

\maketitle

\begin{abstract}
We review novel methods to investigate, control and manipulate
neutral atoms in optical lattices. These setups allow
unprecedented quantum control over large numbers of atoms and thus
are very promising for applications in quantum information
processing. After introducing optical lattices we discuss the
superfluid (SF) and Mott insulating (MI) states of neutral atoms
trapped in such lattices and investigate the SF-MI transition as
recently observed experimentally. In the second part of the paper
we give an overview of proposals for quantum information
processing and show different ways to entangle the trapped atoms,
in particular the usage of cold collisions and Rydberg atoms.
Finally, we also briefly discuss the implementation of quantum
simulators, entanglement enhanced atom interferometers, and ideas
for robust quantum memory in optical lattices.
\end{abstract}

\section{Introduction}

The advent of experimental creation of Bose-Einstein condensates
(BEC) \cite{BECexp} in dilute weakly interacting gases of bosonic
Alkali atoms in 1995 has brought new and exciting prospects to the
fields of atomic physics and quantum state engineering. All the
atoms in a BEC show {\em identical quantum properties} and in many
respects a BEC therefore behaves like one `big' quantum particle.
It also can be imaged without being destroyed, its interactions
with laser light are much stronger than that of a single particle
and a BEC is less fragile than one might expect. Experimentally
observed life times are of the order of several seconds; BEC's can
be moved, shaken and rotated without immediately destroying their
striking quantum features. Therefore they turned out to be an
ideal testing bed for fundamental quantum physics, basic
techniques of quantum state engineering and for investigating
atomic properties \cite{BECrev}. The successful achievement of BEC
and the early studies of their properties led to to the Nobel
Prize in 2001 for the pioneers in this field E.A.~Cornell,
W.~Ketterle, and C.E.~Wieman \cite{nobelwebsite}.

The properties of a BEC described above make it an ideal tool for
cold atom physics and quantum state engineering by quantum optical
methods. One of the most promising experiments for future
applications is the loading of a BEC into an optical lattice
\cite{BEC1Dlatt,optlatt,Greiner}. Optical lattices are periodic
conservative trapping potentials which are created by interference
of travelling laser beams yielding standing laser waves in each
direction (see \fir{fig1}). The laser light induces an AC-Stark
shifts in atoms and thus acts as a conservative periodic
potential. The usage of a BEC for loading has the advantage that
its large atom density allows a filling of few particles per site
$n \gtrsim 1$ while laser cooled atoms loaded into an optical
lattice typically only allow a filling factor smaller than one.
Furthermore the atoms loaded from a BEC are ultracold at
temperatures very close to zero so that they practically behave as
if their temperature was $T = 0$, in particular all of them occupy
the lowest Bloch band. The basic setup of optical lattices and the
loading of a BEC will be described in detail in
Sec.~\ref{BECsetup}.

\begin{figure}[tb]
\begin{center}
\includegraphics[width=10.0cm]{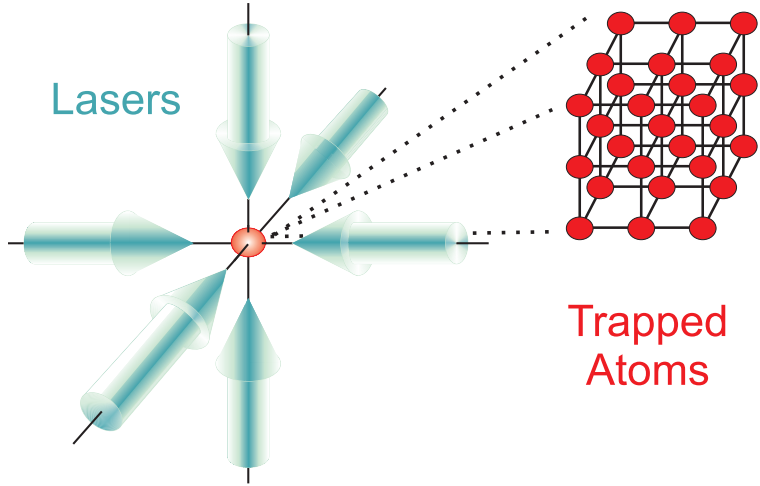}
\caption{Laser setup and resulting optical lattice configuration
in 3D.} \label{fig1}
\end{center}
\end{figure}

An important novel feature which comes about because of the large
filling factor $n \gtrsim 1$ is that interatomic interactions -
due to $s$--wave collisions - of two or more atoms occupying the
same lattice site become important. The interatomic interaction
potential, which usually leads to incoherent collisions in a
thermal cloud of atoms and is responsible for the mean field in a
BEC, causes a coherent energy shift $U$ of two particles occupying
a single lattice site. For the most frequently used alkali atoms
${}^{87}$Rb and ${}^{23}$Na the interaction is repulsive $U>0$ and
thus the interaction energy competes with the kinetic energy $J$
gained by a particle when hopping from one lattice site to the
next. Both of these parameters depend on the laser intensity. When
increasing the laser intensity the barriers between neighbouring
sites increases in height and $J$ goes down. At the same time two
particles sitting in one site are pushed together closer resulting
in an increased interaction energy. Therefore the competition
between kinetic and interaction energy can be controlled by the
laser intensity which is an external easily changeable control
parameter. In Sec.~\ref{MISF} we will describe the physical
consequences of different ratios $J/U$ and investigate the
properties of the resulting ground states.

It turns out that the nature of the resulting quantum ground
states allows for a number of applications ranging from the
controlled creation of molecules \cite{molcre}, the study of
strongly correlated quantum states \cite{Greiner}, enhanced atom
interferometry experiments and precision measurements
\cite{Dorner}, controlled entanglement creation, to quantum
computing \cite{jaksch99,Brennen,jaksch00,Knight,Lewenstein}. In
addition to having well defined initial motional states a number
of different internal (hyperfine) atomic states can be trapped
which results in additional well controlled degrees of freedom.
Optical lattices are very versatile, easily altered by changing
the laser setup and the control over the quantum state of the
atoms by quantum optical techniques is unprecedented. Furthermore,
decoherence times in optical lattices are large compared to
typical experimental times which allows the study of Hamiltonian
dynamics of strongly correlated quantum states with time dependent
parameters and enables quantum information processing. All of
these features make optical lattices a unique system to study. We
will discuss some of the possible applications in Sec.~\ref{App1}
focussing on applications in quantum information processing. We
will show how to implement quantum memory and quantum gates in
optical lattices and review ideas for using optical lattices as a
quantum simulator for complex strongly correlated systems.

The high degree of precision that can be obtained in manipulating
quantum optical systems and the large degree of isolation from the
environment has led to a number of different quantum optical
setups which might be suitable for quantum computing. Several
experiments in the groups of D.~Wineland \cite{IonTrap} and
R.~Blatt \cite{IBKexp} on ion traps have convincingly demonstrated
two qubit gates. The next major step in these efforts is to scale
them to a larger number of ions which will probably be on the
order of $\approx 30$ \cite{IonTrapscale}. Also, photons in an
optical cavity might be used to implement a quantum computer
\cite{Pelli}. The biggest advantage of cavity systems is that they
might allow to couple resting qubits (atoms) to flying qubits
(photons) and thus enable the implementation of a quantum network.

There are also several non-quantum optical systems that have been
identified as candidates for implementing a quantum computer. At
the moment the most advanced system is NMR (Nuclear Magnetic
Resonance) \cite{NMRExp}. However, although NMR systems can be
used to perform demonstrations of quantum computing algorithms on
very small systems it is as yet not known how to scale an NMR
quantum computer. Other promising candidates are superconducting
Josephson junction arrays \cite{JosJun}, quantum dots \cite{Qdots}
and somewhat more exotic systems like e.g.~ammonia molecules
confined inside fullerenes \cite{nanotubes}. Which of these setups
will eventually lead to a successful implementation remains to be
seen. However, as we will show for optical lattice setups in the
remainder of this paper a lot of interesting quantum physics is
contained in the above systems. Therefore, even if some of them
will not succeed in performing scalable quantum computations, it
is certainly worth investigating them to learn more about their
astonishing quantum features.

\section{Loading a BEC into an optical lattice}
\label{BECsetup}

In this section we first briefly describe the creation of an
optical lattice; specifically we show how the interaction between
atoms and a laser field can lead to a conservative trapping
potential. Then we investigate how such a lattice can be used to
trap and manipulate neutral atoms from a BEC. We will also mention
some of the complications and imperfections that arise in this
context but not deal with them in detail.

\subsection{Optical lattices}

\begin{figure}[tb]
\begin{center}
\includegraphics[width=10.0cm]{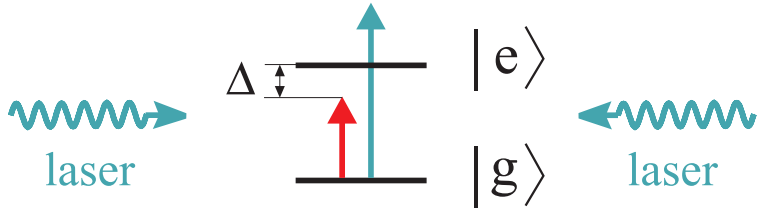}
\caption{Interaction of a laser with a single atom. The red (blue)
arrows indicate red (blue) detuned lasers.} \label{fig2}
\end{center}
\end{figure}

Let us for simplicity consider a two-level atom with an internal
(hyperfine) ground state $\st{g}$ and an excited state $\st{e}$ in
one spatial dimension only coupled to a monochromatic classical
laser field with a detuning $\Delta$ as schematically shown in
\fir{fig2}. The laser is assumed not to cause significant
population in the excited state $\st{e}$ and therefore we may
neglect spontaneous emission from it. In a frame rotating with the
laser frequency the Hamiltonian (with $\hbar=1$) for the atom is
given by
\begin{equation}
H_{\rm atom}= \frac{\hat p^2}{2 m} + \Delta \op{e}{e} -
\frac{\Omega(\hat x)}{2} \left( \op{e}{g} + \op{g}{e} \right).
\end{equation}
Here $\hat x$ and $\hat p$ are the coordinate and momentum
operators respectively, and the mass of the atom is $m$. The laser
drives the transition between the two atomic states with a Rabi
frequency $\Omega$ which is proportional to the laser field and
the dipole transition matrix element of the atom. We now write the
atomic wave function as $\st{\Psi(t)}=\st{\Psi_e(t)} \st{e} +
\st{\Psi_g(t)} \st{g}$ where
$\st{\Psi_{e(g)}(t)}=\skp{e(g)}{\Psi(t)}$ and derive the equations
of motion for the wave functions $\st{\Psi_e(t)}$ and
$\st{\Psi_g(t)}$. We find
 \bea
 i \frac{d \st{\Psi_g(t)}}{dt} &=&
  \frac{\hat p^2}{2 m} \st{\Psi_g(t)} - \frac{\Omega(\hat x)}{2}
  \st{\Psi_e(t)}, \label{wvfung}  \\
 i \frac{d \st{\Psi_e(t)}}{dt} &=&
  \left(\frac{\hat p^2}{2 m} + \Delta \right) \st{\Psi_e(t)} - \frac{\Omega(\hat x)}{2}
  \st{\Psi_g(t)} \label{wvfune} .
 \eea
If the laser is far detuned, i.e.~$|\Delta| \gg |\Omega|$ it is
permissible to adiabatically eliminate the excited internal state
of the atom $\st{e}$. The adiabatic elimination consists in
assuming that the wave function $\st{\Psi_e(t)}$ adiabatically
follows any change in the ground state wave function
$\st{\Psi_g(t)}$. We can thus set the left hand side of
\eqr{wvfune} equal to zero. Furthermore we assume the kinetic
energy of an excited atom (and thus also the recoil energy
$E_R=k^2/2m$ where $k=2 \pi/\lambda$ with $\lambda$ the laser wave
length) to be much smaller than the detuning $E_R \ll |\Delta|$
and neglect it in \eqr{wvfune} compared to $\Delta$. The equation
\eqr{wvfune} is now easily solvable and substituting its solution
into \eqr{wvfung} we find that the atom now moves according to the
Schr{\"o}dinger equation
 \be
 i \frac{d \st{\Psi_g(t)}}{dt} =
  \left(\frac{\hat p^2}{2 m} + V_0(\hat x) \right) \st{\Psi_g(t)}, \label{exfun}
 \ee
where we have defined the optical potential as
\begin{equation}
V_0(x)=-\frac{\Omega(x)^2}{4 \Delta}.
\end{equation}
We will always assume the atom to be interacting with a standing
light wave which arises from the interference of two counter
propagating laser beams. This means that the spatial dependence of
the Rabi frequency is given by
\begin{equation}
\Omega(x)= \Omega_0 \sin(k x),
\end{equation}
where $\Omega_0$ is a constant depending on the laser intensity
and the properties of the atom. We therefore find that the motion
of the atom is effectively described by a periodic conservative
potential with period $a=\lambda/2$ and a depth controlled by the
laser intensity. For a sufficiently deep lattice the atoms can be
trapped near the potential minima which are also called lattice
sites. In the vicinity of each lattice site the trap potential is
well approximated by a harmonic oscillator potential with
frequency
 \be
 \omega_T=\left|\frac{\Omega_0 k}{\sqrt{2 \Delta m}}\right|.
 \ee

\subsection{Imperfections}

Above we have totally neglected spontaneous emission from the
excited atomic level. We give an estimate of this effect by
assuming that the atom is localized close to the minima of the
optical lattice potential. For blue detuned optical lattices
($\Delta <0$) the lattice sites are at positions of zero laser
intensity while for a red detuned optical lattice ($\Delta > 0$)
they are at locations of maximum laser intensity. Therefore in a
blue lattice the effects of spontaneous emission on trapped
particles are less strong than for red lattices. In a blue lattice
we find the effective spontaneous emission rate $\gamma_{\rm
eff}^b$ of a trapped atom
\begin{equation}
\gamma_{\rm eff}^b \approx -\frac{\gamma}{4 \Delta} \omega_T,
\end{equation}
where $\gamma$ is the spontaneous emission rate from the excited
atomic level. Therefore spontaneous emission will be negligible
for large detunings $\Delta$. Indeed, it has been shown in an
experiment by Friebel et al. \cite{BHMFriebel} that the
spontaneous emission rate in far detuned optical lattices can be
of the order of several minutes.

\begin{figure}[tb]
\begin{center}
\includegraphics[width=10.0cm]{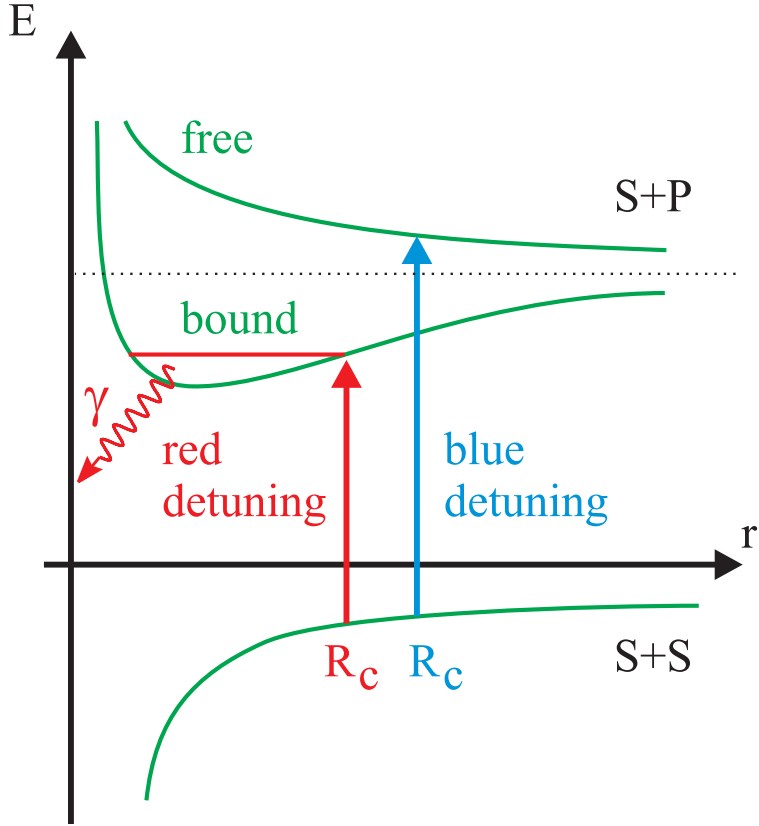}
\caption{Schematic interaction potential curves for two atoms at
distance $r$. The spontaneous emission from a bound molecular
state $\gamma$ is also indicated.} \label{fig3}
\end{center}
\end{figure}

Since we are interested in having a few atoms per lattice we have
to consider the effects of light induced inelastic collisions
between two atoms. In \fir{fig3} we have schematically shown the
interaction potentials for the different internal states $(S+S)$
and $(P+S)$ of two atoms approaching each other at a distance $r$.
As a consequence of this interaction the detuning $\Delta$ of the
laser field changes as the particles come close to each other and
becomes $\Delta=0$ at the so called Condon point $R_C$. The
probability for the two particles to be excited to the states
$P+S$ increases around $R_C$. The atoms can then either form a
quasistable molecule (for red detuning) which decays by
spontaneously emitting a photon or they can be promoted to an
unbound $S+P$ state with a large kinetic energy equal to the
detuning of the laser at $r=\infty$. Both of these effects should
be kept as small as possible to minimize loss from the optical
lattice. There are several ways to do so:
\begin{itemize}

\item Avoid a Condon point by detuning far to the red. However,
this will lead to a larger spontaneous emission rate compared to
the case of a blue lattice.

\item Avoid a red detuning that puts the Condon point close to a
quasi bound molecular state (shown in \fir{fig3}) as this enhances
the probability of creating such molecules.

\item Choose the detuning such that the two potential curves have
the largest possible angle with each other at the Condon point.
This makes the probability of finding the two particles in the
region where $\Delta \approx 0$ small.

\end{itemize}
The above conditions for neglecting spontaneous emission and
making laser induced collisions unimportant can be fulfilled quite
well experimentally and then our above simple model for an optical
lattice as a conservative potential is valid.

\subsection{Experimental setups}

The situation where one pair of counter propagating laser beams
interferes and yields a trap potential in 1D while the atoms are
only weakly trapped by a magnetic field (or not trapped at all) in
the other two dimensions is usually called a 1D optical lattice. A
BEC was first loaded into a 1D lattice by B.P.~Anderson and
M.A.~Kasevich \cite{BEC1Dlatt}. In this experiment a magnetically
trapped BEC is loaded into a weak vertical optical lattice while
at the same time turning off the magnetic trapping almost
completely. The experiment is designed such that the interplay
between the gravitational force and the optical lattice leads to a
peculiar interference pattern of atoms falling out of the lattice
which allows to study the coherence properties of the original
BEC.

A 1D optical lattice setup is theoretically easily extended to the
3D situation as shown in \fir{fig1}. Three pairs of counter
propagating laser beams interfere to produce an optical potential
which traps the atoms in each dimension. Experimentally, however,
the achievement of this situation is less straightforward. The
experimental setup has to allow for optical access to the BEC from
three different orthogonal directions and this is usually
incompatible with producing the magnetic trap required to create
the BEC. For this reason the group headed by T.~H{\"a}nsch in
Munich realized the concept of spatially separating the production
of a BEC and performing the actual optical lattice experiment.
Their setup is schematically shown in \fir{fig4}. A number of
coils are used to create an inhomogeneous magnetic field that
guides the BEC out of its original position and transports it into
the `science chamber' where optical access from three directions
is possible and allows the realization of a 3D optical lattice.
Next we will discuss the novel features of a BEC loaded into such
a 3D optical lattice. In particular we will investigate the
effects of interactions between the particles which are
responsible for striking new features in comparison to previous
experiments on optical lattices.

\begin{figure}[tb]
\begin{center}
\includegraphics[width=10.0cm]{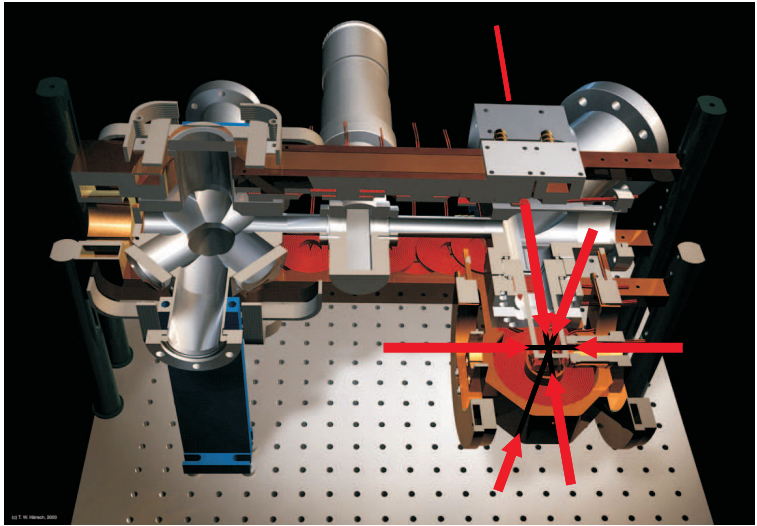}
\caption{Experimental setup used in Munich to load a BEC into a 3D
optical lattice \cite{Greiner}. The BEC is produced in a magnetic
trap located at the left hand side of the figure and is then
transported to the right where optical access from all directions
is possible as shown by the red arrows.} \label{fig4}
\end{center}
\end{figure}

\section{Superfluids and Mott insulators}
\label{MISF}

The above setup with 3D lattice geometry was predicted to yield a
qualitatively new behaviour of a dense cloud of ultracold atoms in
1998 \cite{jaksch98} and was experimentally realized in 2001 by M.
Greiner et al.~\cite{Greiner}. We can gain some insight into the
properties of the atoms trapped in the lattice by looking at two
limiting cases; the superfluid limit where the kinetic energy
dominates the repulsive interaction and the opposite limit which
is called the Mott insulating limit.

\subsection{Superfluid limit}

When the optical lattice is shallow the kinetic energy dominates
over the repulsion of two particles sitting in the same lattice
site. In this situation nearly all the atoms occupy the same
single particle state. For very small interaction energies this
single particle state is very close to the Bloch wave function
with quasi momentum $q=0$ of the lowest Bloch band, i.e.~the
single particle ground state of the periodic potential. As this
wave function is delocalized over the system there is long range
off-diagonal coherence over the whole lattice. States like this
are very well described by the Gross-Pitaevskii equation
\cite{BECrev} which takes the interaction into account via a
simple mean field approximation. Neglecting interactions
altogether we can understand the form of the underlying
Hamiltonian intuitively. We define a bosonic destruction operator
$a_l$ destroying an atom in the lattice site labelled $l$. Its
adjoint operator $a_l^\dagger$ correspondingly creates a particle
in this lattice site and $n_l=a_l^\dagger a_l$ counts the number
of particles in site $l$. By hopping (tunneling) from one lattice
site to an adjacent site an atom loses a kinetic energy $J$
depending on the height of the barrier between the sites and the
Hamiltonian therefore reads
 \be
 H_{\rm SF}=-J \sum_{\langle l,m \rangle} a_l^\dagger a_m,
 \label{eqSF}
 \ee
where the sum runs over all pairs of nearest neighbours in the
lattice. An atom in the lattice will therefore minimize its energy
when it is in a superposition state of many lattice sites.

A more detailed investigation shows that the state
 \be
 \st{\Psi_{\rm SF}} \propto \left( \sum_l a^\dagger_l \right)^N
 \st{\rm vac}
 \ee
of $N$ particles with $\st{\rm vac}$ the vacuum minimizes the
energy. This state clearly corresponds to each particle occupying
the lowest Bloch state and spreads the wave function of each
particle over the whole lattice as indicated in \fir{fig5}a. The
state therefore shows large off diagonal long range coherence as
can be seen by correlating the destruction of an atom in lattice
site $l$ with the creation of a particle in site $m$. The
corresponding correlation function is found by the expectation
value $\rho_{l,m}=\langle a_l^\dagger a_m \rangle$. In the SF
regime this correlation function is independent of the distance
between sites $l$ and $m$ and its modulus is given by
$|\rho_{l,m}|=N/M$ where $M$ is the total number of lattice sites.
We also note that the particle number fluctuations $(\Delta
n_l)^2=N/M$ in this state are large.

The lowest lying excitations in the SF limit are determined by the
Bloch band and consist of waves with quasi momentum $q$. The
energy of the first excited state scales like $\epsilon_1 \propto
J/M^2$ which tends to $0$ for large systems $M\rightarrow \infty$.
This vanishing gap between the ground state and low lying excited
states makes the SF state rather fragile and seems to render it
very difficult to adiabatically load a BEC into an optical
lattice. Fortunately, however, the turning on of a lattice does
not induce quasi-momenta and therefore does not excite low lying
excitations. Therefore loading a lattice from a BEC is possible at
reasonable speed without significant heating of the atomic cloud.

\begin{figure}[tb]
\begin{center}
\includegraphics[width=10.0cm]{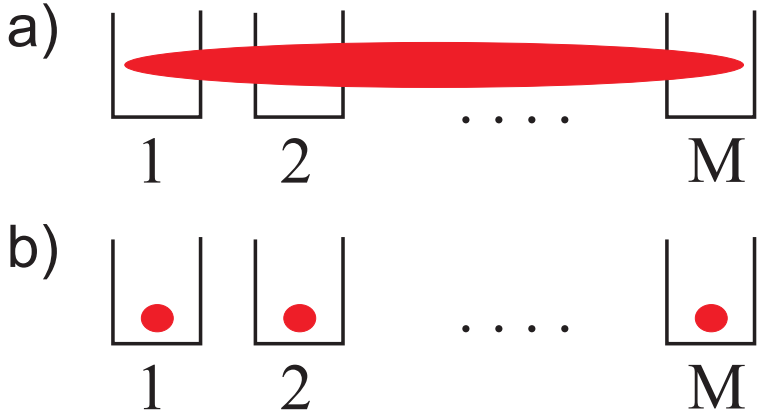}
\caption{Schematic representation of a) a SF state and b) a MI
state with commensurate filling $n=1$ in an optical lattice with
$M$ sites.} \label{fig5}
\end{center}
\end{figure}

\subsection{Mott insulating limit}

By increasing the lattice depth the kinetic energy goes down
because the barrier height increases. At the same time the size of
each lattice site decreases which causes the atoms to repel each
other more strongly. Again we can understand this situation
intuitively. We neglect any kinetic energy and only consider the
interaction energy of two particles in the same lattice site. As
usual the two particle interaction energy increases with the
square of the particle density which itself is proportional to the
square of the number of particles per well. The Hamiltonian
therefore is given by
 \be
 H_{\rm MI}= \frac{U}{2} \sum_{l} n_l (n_l-1).
 \label{eqMI}
 \ee
Here the parameter $U$ is the interaction energy of two particles
in the same lattice site. Note the $-1$ term which is usually
neglected in most mean field theories. It ensures that a single
particle does not gain any interaction energy.

In this situation particles in the optical lattice minimize their
energy by staying away from each other instead of spreading out
over the whole lattice. Let us consider the case of commensurate
filling $M=N$ where the smallest interaction energy is obtained by
putting each particle into its individual lattice site. A state
like this is written as
 \be
 \st{\Psi_{\rm MI}} \propto \prod_l a^\dagger_l \st{\rm vac},
 \ee
is called a Mott insulator, and schematically shown in
\fir{fig5}b. It does not have any interaction energy
$\ME{\Psi_{\rm MI}}{H_{\rm int}}{\Psi_{\rm MI}}=0$ while the above
SF state would have an interaction energy of $\ME{\Psi_{\rm
SF}}{H_{\rm int}}{\Psi_{\rm SF}} = U N (N-1)/2 M^2$. Furthermore,
the MI state has no long range coherence $\rho_{l,m}=0$ for $l
\neq m$ and also the particle number fluctuations are zero $\Delta
n_l^2=0$.

The MI state is much more stable than the SF state. The lowest
lying excitations are obtained by taking one particle out of its
site and putting it on top of an atom sitting in another lattice
site. The energy required for this is $U$ independent of the size
$M$, i.e.~the system is characterized by a finite gap $U$ large
compared to the small excitation energies in the SF limit.

Experimentally the situation of commensurate filling can be
achieved by suitably superimposing the optical lattice with a much
weaker magnetic trapping potential. The atoms fill those lattice
sites which are in the center of the wide trap and avoid the outer
ones so that we obtain commensurate filling in the center of the
lattice. The magnetic trap acts as a `soft' boundary which adjusts
to the number of particles. Note, however, that the nature of the
MI will be altered close to this boundary.

\subsection{Intermediate regime}

If the two parameters $J$ and $U$ are of similar magnitude the
situation is more complicated. Kinetic energy and interactions
compete with each other and only a detailed investigation shows
whether the system is a SF or a MI. For commensurate filling of
the lattice with a filling factor of $n=1$ one easily finds that
the interaction wins for $U>U_c$ while the kinetic energy
dominates for $U<U_c$, where the critical point is $U_c \approx
5.8 z J$. Here $z=2d$ is the number of nearest neighbours in a
$d$-dimensional lattice. For larger commensurate fillings of the
lattice the value of $U_c$ increases because the interaction
energy goes up quadratically with the filling while the gained
kinetic energy only increases linearly with the filling factor.

\begin{figure}[tb]
\begin{center}
\includegraphics[width=10.0cm]{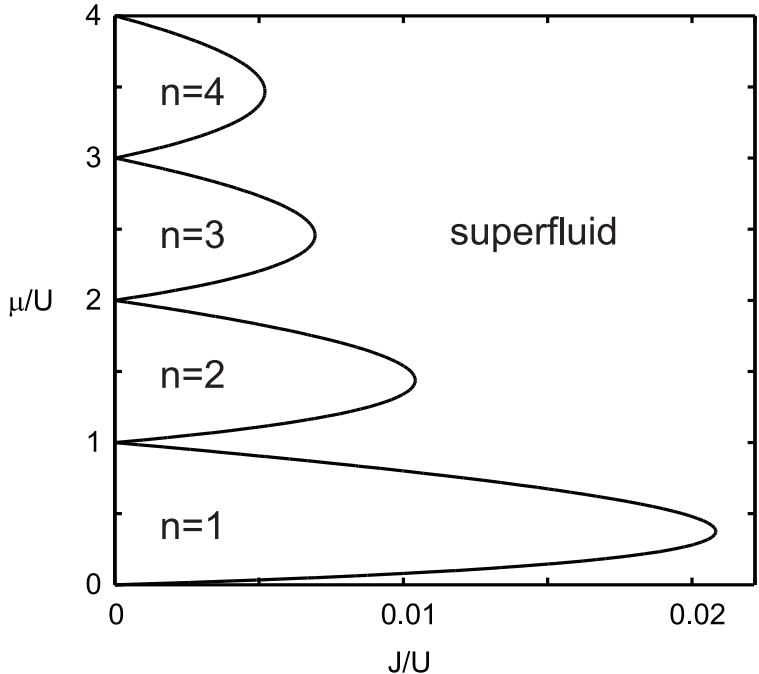}
\caption{Phase diagram of the 3D Bose-Hubbard model showing the MI
regions with filling factor $n$ and the SF region for different
ratios $J/U$ and chemical potentials $\mu/U$. Note that the simple
mean field approximation used here does not yield $U_c=5.8 z J$.}
\label{fig6}
\end{center}
\end{figure}

For incommensurate fillings the system is best described in the
grand canonical ensemble where the mean particle number is fixed
by a particle reservoir of chemical potential $\mu$. By using
simple mean field theory the SF and MI regions can be found
approximately; the numerical result is shown in \fir{fig6}. When
varying $J$ over the critical point the ground state abruptly
changes its nature and the transition between the SF and the MI
regime at temperature $T=0$ takes place in a so called quantum
phase transition if the lattice is sufficiently large. We can
understand the basic physics of a quantum phase transition by
considering a Hamiltonian of the form $H=H_1 + g H_2$ where the
two parts of $H$ commute, i.e.~$[H_1,H_2]=0$ and $g>0$ is a
parameter. Since $H_1$ and $H_2$ commute the eigenstates of $H$
are independent of $g$ while the eigenvalues change with $g$. For
small values of $g<g_c$ the ground state of $H$ will be determined
by the ground state of $H_1$ while for large values of $g>g_c$ it
is given by the ground state of $H_2$. If these two ground states
have totally different properties the system alters its behaviour
suddenly when the parameter $g$ is changed over the critical point
$g_c$. A quantum phase transition also occurs when the commutator
of the two parts of the Hamiltonian is not exactly zero but does
not increase with the size of the system and therefore can be
neglected in the case of large systems. This is exactly the
situation we encounter in the so called Bose-Hubbard model with
Hamiltonian $H_{\rm BHM} =  H_{\rm SF} +  H_{\rm MI}$ we are
considering here for a changing ratio of $J/U$.

An optical lattice has the unique feature of permitting studies of
the dynamics of the quantum phase transition in $H_{\rm BHM}$,
i.e., the parameters $U$ and $J$ can be changed on a time scale
much smaller than the coherence time of the system but still
sufficiently slow to avoid significant excitations. The
experiments carried out by M.~Greiner et al.~\cite{Greiner} to
study this dynamics started out from a Bose-condensate in a
magnetic trap. Then the optical lattice was (adiabatically) turned
on. Its depth determines the ratio $J/U$ and thus the kind of
ground state in the optical lattice. To measure the properties of
the atomic state the lattice potential is then turned off and the
particles are allowed to expand freely before their density
distribution is measured. In the case of a SF state which has long
range off diagonal coherence the particles interfere and the
resulting interference pattern clearly indicates the band
structure of the lattice. When the system is in a MI state there
is no long range coherence and therefore the interference pattern
disappears. This behaviour is similar to interference in optics
experiments where interference patterns also vanish when there is
lack of coherence. The corresponding interference patterns as
measured in \cite{Greiner} are shown in \fir{fig7} for different
depths of the optical lattice.

\begin{figure}[tb]
\begin{center}
\includegraphics[width=10.0cm]{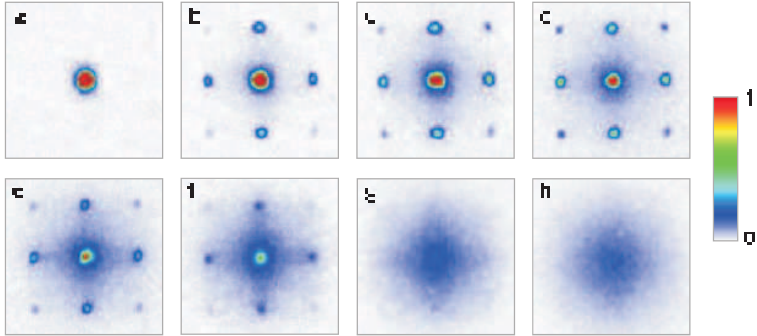}
\caption{Interference patterns for decreasing $J/U$ (a-h). For
$U<U_c$ the long range coherence leads to an interference pattern
which clearly indicates the band structure (a-d). The interference
fringes vanish when the system enters the MI regime $U>U_c$
(e-h).} \label{fig7}
\end{center}
\end{figure}

Next we discuss some of the applications of MI states in optical
lattices focussing on applications in quantum computing.
Afterwards we look at atomic interferometry enhanced by using
multi particle entangled states.

\section{Quantum information processing in optical lattices}
\label{App1}

The experiments on the SF to MI transition are a very important
step towards realizing theoretical ideas for controlled
entanglement creation in optical lattices by interactions between
neutral atoms \cite{jaksch99,jaksch00,Brennen,Knight}. Together
with single particle manipulations such entanglement operations
constitute the basic building blocks for a quantum computer
\cite{SetOfGates} which fulfils all the basic requirements as set
out by D.~DiVincenzo \cite{DiVincenzo}. First important
experimental steps towards realizing multi particle entangled
states have recently been achieved by Mandel {\em et al.}
\cite{BlochEnt}. In this section we will first discuss the
requirements for implementing a quantum computer and then show how
they can be satisfied in an optical lattice.

\subsection{What is quantum information processing?}

A computation can be considered as a physical process that
transforms an input into an output. A classical computation is
that in which the physical process is based on classical laws
while the physical process in quantum computation is based on
quantum laws and in particular on the superposition principle. In
quantum information processing the inputs and outputs are encoded
into the states of the system and the computation evolves the
input state by a designed well controlled unitary time evolution.
In general, if our algorithm consists of evaluating a given
function $f$, we can design an interaction Hamiltonian such that
the resulting unitary evolution transforms the input states
according to $|j\rangle \otimes |0\rangle \rightarrow |j\rangle
\otimes \st{f(j)}$ ($j=1,N$).

Using a quantum computer we can do more than with a classical
computer. When preparing the input state in a superposition of all
$L=2^N$ basis states for $N$ qubits it will be transformed
according to
 \be
\frac{1}{\sqrt{L}}\sum_{\ell=1}^{L}|\ell\rangle \otimes |0\rangle
\rightarrow \frac{1}{\sqrt{L}}\sum_{\ell=1}^{L}|\ell \rangle
\otimes |f(\ell)\rangle
 \ee
in a single run. All the values of the function $f$ are present in
this superposition. However, we do not have access to all of this
information since after a measurement we will only obtain one
result with a certain probability. The property of using quantum
superpositions to run the computer only once was termed quantum
parallelism by Feynman.

\subsubsection{Requirements}

A quantum computer consists of a quantum register (in our case an
array of atoms in the optical lattice) that can be manipulated and
measured in a controlled way. If we want to build a quantum
computer, we need the following elements \cite{DiVincenzo}:

(i) A set of qubits which are two-level systems forming the
quantum register. In optical lattices this set of qubits is formed
by two internal states of the atoms trapped in the lattice. We
denote by $\{\st{0}_j,\st{1}_j\}$ two orthogonal states of the
$j$-th qubit (also called the computational basis), so that a
state $\st{\Psi}$ of all the qubits in the quantum register can be
written as
 \bea
 \st{\Psi} &=& \sum_{j_{1},j_{2},\ldots
 j_{N}=0}^{1}c_{j_{1}j_{2}\ldots j_{N}}\st{j_1}_1 \otimes
 \st{j_2}_2 \otimes \ldots \otimes \st{j_N}_N \nonumber \\
 &\equiv& \sum_{j_{1},j_{2},\ldots
 j_{N}=0}^{1}c_{j_{1}j_{2}\ldots j_{N}}\st{j_1,j_2,\ldots j_N}.
 \eea
These qubits can be in superposition and entangled states, which
gives the extraordinary power to the quantum computer; they also
have to be well isolated from the environment to prevent
decoherence.

(ii) A universal set of quantum gates which allows the controlled
manipulation of the qubits according to any unitary operation $U$
so that $|\Psi \rangle \rightarrow U|\Psi \rangle .$ Fortunately,
this task is enormously simplified given the fact that any $U$ can
be decomposed as a product of gates belonging to a small set, a
so-called universal set of gates, i.e.~if we are able to perform
the gates of this set we will be able to perform any unitary
operation on the register by applying a sequence of them. There
are many sets of universal gates and we will concentrate on the
set containing one two qubit phase gate and single qubit
operations as this is the one most naturally arising in optical
lattice setups. Note that two-qubit gates require interactions
between the qubits, and therefore are the more difficult ones in
practice.

(iii) Detection of the output state. One should be able to measure
each qubit in its computational basis. This process requires the
interaction with a measurement apparatus in an irreversible way
and is thus not governed by a unitary time evolution.

(iv) Initial state preparation. We must be able to erase quantum
registers to prepare the initial quantum state, for example the
state $|0,0,\ldots ,0\rangle $. This is actually not an extra
requirement, it is enough to detect the qubits and to apply a
single-qubit gate flipping the qubit $\st{1} \rightarrow \st{0}$
and $\st{0} \rightarrow \st{1}$ if necessary.

(v) Scalability of the system. The difficulty of performing gates,
measurements, etc., should not grow (exponentially) with the
number of qubits. Otherwise, the gain in the quantum algorithms
would be lost.

(vi) In addition to the above we also require networking ability.
The static qubits used to perform quantum computations should be
transformable into flying qubits which can easily be transmitted
between different locations.

In the following we will discuss how to implement an initialized
quantum register and a universal set of quantum gates in an
optical lattice setup. Due to its periodic structure the optical
lattice provides an intrinsically scalable setup. The problem of
detecting the output state is closely related to addressing single
qubits as required for performing single qubit operations. A
discussion of how to achieve networking ability lies beyond the
scope of this paper.

\subsection{Quantum memory and single qubit gates in optical
lattices} \label{Qmemsinqub}

In optical lattices two states of the ground state hyperfine
manifold of the trapped atoms $\st{a} \equiv \st{0}$ and $\st{b}
\equiv \st{1}$ are ideal candidates to implement a qubit. However,
in order to be able to manipulate and control atomic qubits it is
necessary to know the position of each of the atoms precisely.
While random filling of optical lattices from laser cooled atoms
and a superfluid filling of a lattice (with large particle number
fluctuations as discussed above) do not provide sufficient
knowledge on the position of the atoms a MI state is ideally
suited for this purpose. All the lattice sites are filled,
i.e.~each lattice site contains a qubit and the fluctuations in
the occupation numbers are very small. For an appropriate choice
of internal atomic states the optical lattice allows to trap both
of them and thus can hold a quantum register for storing quantum
information.

In principle it is straightforward to induce single qubit gates by
using Raman transitions between the two internal states $\st{a}$
and $\st{b}$. Raman transitions with Rabi frequency $\Omega_R$ and
detuning $\delta$ are described by the Hamiltonian
 \be
 H_R=\frac{1}{2}\left(\Omega_R\op{a}{b} + h.c. \right) + \delta \op{b}{b},
 \ee
which induce rotations of the qubit state on the Bloch sphere. The
axis and angle of this rotation depend on the choice of laser
parameters and can be chosen freely. The major problem in inducing
single qubit operations is the addressing of a single atom as it
is difficult to focus a laser to spots of order of an optical wave
length which is the typical separation between atoms in the
lattice. Possible solutions to this difficulty are using schemes
for pattern loading \cite{jaksch98, Phillips2003}, where only
specific lattice sites are filled with atoms (cf.~\fir{fig8}) or
using additional marker atoms which specify the atom the laser is
supposed to interact with.

\begin{figure}[tb]
\begin{center}
\includegraphics[width=8.0cm]{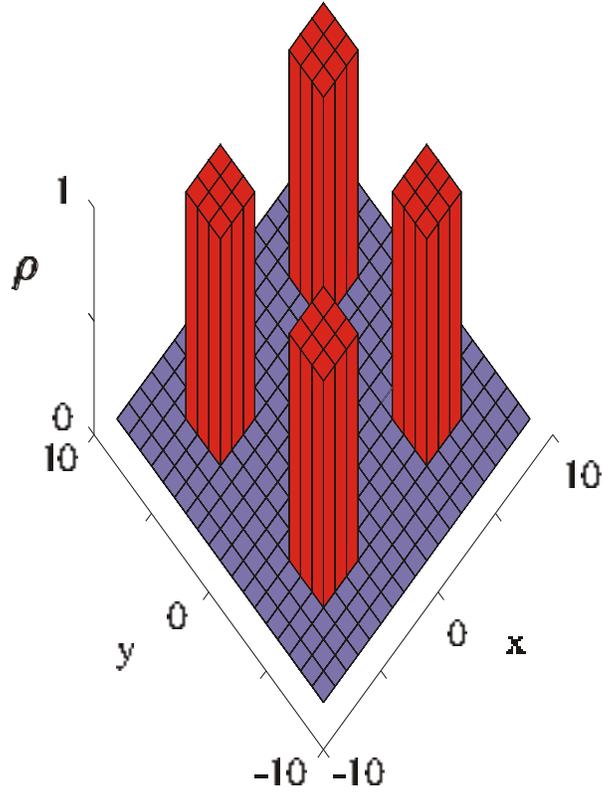}
\caption{Selective filling of optical lattice sites in a
superlattice structure. The red boxes indicate lattice sites
filled with exactly one particle. The blue boxes show lattices
sites that are empty.} \label{fig8}
\end{center}
\end{figure}

MI atoms in an optical lattice have already experimentally been
used as qubits and it has been shown that they support multi
particle entangled states \cite{BlochEnt}. The robustness of these
qubits is, however, limited by stray magnetic fields and spin echo
techniques need to be used to perform the experiments
successfully. A different approach to obtaining robust quantum
memory uses a more sophisticated encoding of the qubits
\cite{Dorner}. A 1D chain with an even number of atoms encodes a
single qubit in the states $\st{0}=\st{abab \cdots ab}$ and
$\st{1}=\st{baba \cdots ba}$. These states both contain the same
number of atoms in internal states $\st{a}$ and $\st{b}$ and
therefore interact with magnetic fields in exactly the same way
avoiding dephasing of the quantum information stored in the chain.
This and the fact that all the atoms have to flip their internal
state to get from one logical state to the other make these qubits
very robust against the most dominant experimental sources of
decoherence. However, at the same time this robustness makes it
more difficult to manipulate them. A laser pulse no longer
corresponds to a simple rotation on the Bloch sphere spanned by
the logical states $\st{0}$ and $\st{1}$ which makes the
realization of a single qubit gate difficult. We will describe how
these problems can be circumvented in Sec.~\ref{App2} and further
details can be found in \cite{Dorner}.

\subsection{Two qubit operations}

Implementing a two qubit gate is more challenging than the single
qubit gates. The different schemes for two qubit gates can be
classified in two categories. The first version relies on the
concept of a quantum data bus; the qubits are coupled to a
collective auxiliary quantum mode, like e.g.~a phonon mode in an
ion trap, and entanglement is achieved by swapping the qubits to
excitations of the collective mode. The second concept which is
the basis for two qubit gates between atoms in optical lattices
deploys controllable internal-state dependent two-body
interactions. Examples for different interactions are coherent
cold collisions of atoms, optical dipole-dipole interactions
\cite{jaksch99,Brennen} and the `fast' two-qubit gate based on
large permanent dipole interactions between laser excited Rydberg
atoms in static electric fields \cite{jaksch00}. Besides these
dynamical schemes for entanglement creation it is also possible to
generate entanglement by purely geometrical means \cite{Duan01}.
We will now discuss the different ways to achieve two qubit
operations in optical lattices

\subsubsection{Entanglement via coherent ground state collisions}

The interaction terms introduced in \eqr{eqMI} which describe
$s$--wave collisions between ultracold atoms in one lattice site
are analogous to Kerr nonlinearities between photons in quantum
optics. For atoms stored in optical lattices these nonlinear
atom-atom interactions can be large \cite{jaksch99}, even for
interactions between individual pairs of atoms, thus providing the
necessary ingredients to implement two-qubit gates.

\begin{figure}[tb]
\begin{center}
\includegraphics[width=10.0cm]{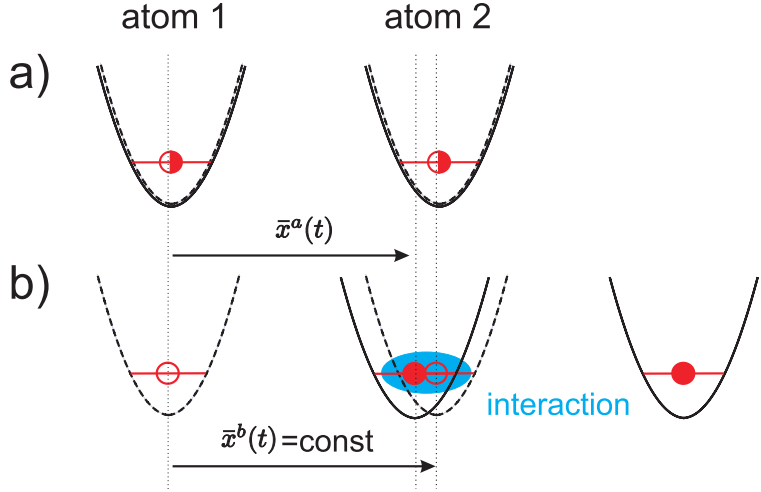}
\caption{We collide one atom in internal state $\st{a}$ (filled
circle, potential indicated by solid curve) with a second atom in
state $\st{b}$ (open circle, potential indicated by dashed
curve)). In the collision the wave function accumulates a phase
according to Eq.~(\protect\ref{transf}). a) Configurations at
times $t=\pm \tau$ and b) at time $t$.} \label{fig9}
\end{center}
\end{figure}

We consider a situation where two atoms in a superposition of
internal states $\st{a}$ and $\st{b}$ are trapped in the ground
states of two optical lattice sites (see \fir{fig9}a). Initially,
at time $t=-\tau$ these wells are centered at positions
sufficiently far apart so that the particles do not interact. The
optical lattice potential is then moved state selectively and for
simplicity we assume that only the potential for a particle in
internal state $\st{a}$ moves to the right and drags along an atom
in state $\st{a}$ while a particle in state $\st{b}$ remains at
rest. Thus the wave function of each atom splits up in space
according to the internal superposition of states $\st{a}$ and
$\st{b}$. When the wave function of the left atom in state
$\st{a}$ reaches the second atom in state $\st{b}$ as shown in
\fir{fig9}b they will interact with each other. However, any other
combination of internal states will not interact and therefore
this collision is conditional on the internal state. A specific
laser configuration achieving this state dependent atom transport
has been analyzed in Ref.~\cite{jaksch99} for Alkali atoms, based
on tuning the laser between the fine structure excited $p$-states.
The trapping potentials can be moved by changing the laser
parameters. Such trapping potentials could also be realized with
magnetic and electric microtraps \cite{Calarco00}.

We therefore only need to consider the situation where atom $1$ is
in state $\st{a}$ and particle $2$ is in state $\st{b}$ to analyze
the interactions between the two atoms. The positions of the
potentials are moved along trajectories $\bar x^{a}(t)$ and $\bar
x^{b}(t)={\rm const}$ so that the wave packets of the atoms
overlap for a certain time, until they are finally restored to the
initial position at the final time $t=\tau$. This situation is
described by the Hamiltonian
\begin{equation}
H=\sum_{\beta =a,b}\left[ \frac{(\hat{p}^{\beta
})^{2}}{2m}+V^{\beta
}\left( \hat{x}^{\beta }-\bar x^{\beta }(t)\right) \right] +u^{\mathrm{%
ab}}(\hat{x}^{a}-\hat{x}^{b}).
\end{equation}
Here, $\hat{x}^{a,b}$ and $\hat{p}^{a,b}$ are position and
momentum operators, $V^{a,b}\left( \hat{x}^{a,b}- \bar x^{a,b}(t)
\right)$ describe the displaced trap potentials and $u^{\rm{ab}}$
is the atom--atom interaction term (which lead to the interaction
term in \eqr{eqMI}). Ideally, we want to implement the
transformation from before to after the collision,
\begin{equation}
\psi _{0}^{a}(x^{a}-\bar x^{a}(-\tau))\psi _{0}^{b}(x^{b}-\bar
x^{b}(-\tau))\rightarrow e^{i\phi }\psi _{0}^{a}(x^{a}-\bar
x^{a}(\tau))\psi _{0}^{b}(x^{b}-\bar x^{b}(\tau)), \label{transf}
\end{equation}
where each atom remains in the ground state $\psi_{0}^{a,b}$ of
its trapping potential and preserves its internal state. The phase
$\phi =\phi^{a}+\phi^{b}+\phi^{\rm{ab}}$ will contain a
contribution $\phi^{\rm{ab}}$ from the interaction (collision) and
(trivial) single particle kinematic phases $\phi^{a}$ and
$\phi^{b}$. The transformation \eqr{transf} can be realized in the
\emph{adiabatic limit}, whereby we move the potentials slowly on
the scale given by the trap frequency, so that the atoms remain in
their motional ground state. In this case the collisional phase
shift is given by $\phi ^{\rm{ab}}=\int_{-\infty }^{\infty
}dt\Delta E(t)/\hbar$, where $\Delta E(t)$ is the energy shift
induced by the atom--atom interactions
\begin{equation}
\Delta E(t)=\frac{4\pi a_{s}\hbar ^{2}}{m}\int dx|\psi
_{0}^{a}\left( x- \bar x^{a}(t)\right) |^{2}|\psi _{0}^{b}\left(
x-\bar x^{b}(t)\right) |^{2}, \label{deltaE}
\end{equation}
with $a_{s}$ the $s$--wave scattering length. In addition we
assume that $|\Delta E(t)|\ll \hbar \omega_T $ so that no sloshing
motion is excited.

The interaction phase thus only applies when the atoms are in
internal state $\st{a,b}$ but not otherwise. Carrying out the
above state selective collision with a phase $\phi ^{\rm{ab}}=\pi$
we obtain (up to trivial phases) the mapping (as above we identify
$\st{a} \equiv \st{0}$ and $\st{b} \equiv \st{1}$)
\begin{eqnarray}
|0,0\rangle &\rightarrow &|0,0\rangle \nonumber \\
|0,1\rangle &\rightarrow &-|0,1\rangle \nonumber  \\
|1,0\rangle &\rightarrow &|1,0\rangle \nonumber \\
|1,1\rangle &\rightarrow &|1,1\rangle
\end{eqnarray}
which realizes a two-qubit phase gate that is universal in
combination with single-qubit rotations.

\subsubsection{State-selective interaction potential}

An alternative possibility, for a nontrivial logical phase to be
obtained, is to rely on a state-independent trapping potential,
while defining a procedure where different logical states couple
to each other with different energies. An example is given by the
interaction between state-selectively switched electrical dipoles
\cite{jaksch00}.

In each qubit, the hyperfine ground states $\st{a} \equiv \st{0}$
is coupled by a laser to a given Stark eigenstate $|r\rangle$
which does not correspond to a logical state. The internal
dynamics is described by a model Hamiltonian
 \bea
 H_I(t,{\bf x}_1,{\bf x}_2)&=& \sum_{j=1,2} \left[\delta_j(t) |r\rangle_j
 \langle r| - \frac{\Omega_j(t,{\bf x}_j)}{2} \left(|a\rangle_j
 \langle r|+\rm{h.c.} \right)\right]\nonumber\\
 &&\mbox{}+u |r\rangle_1 \langle r|\otimes |r\rangle_2 \langle r| ,
 \label{modelRyd}
 \eea
with $\Omega_j(t,{\bf x}_j)$  Rabi frequencies, and $\delta_j(t)$
detunings of the exciting lasers. Here, $u$ is the dipole-dipole
interaction energy between the two particles. We have neglected
any loss from the excited states $|r\rangle_j$. We discuss two
possible realizations of two qubit gates with this dynamics. The
most straightforward way to implement a two-qubit gate is to just
switch on the dipole-dipole interaction by exciting each qubit to
the auxiliary state $|r\rangle$, conditioned on the initial
logical state. This can be obtained by two resonant
($\delta_1=\delta_2=0$) laser fields of the same intensity,
corresponding to a Rabi frequency $\Omega_1=\Omega_2\gg u$. After
a time $\tau=\varphi/u$, the gate phase $\varphi$ is accumulated
and the particles can be taken again to the initial internal
state. However, besides $\varphi$ being sensitive to the atomic
distance via the energy shift $u$, during the gate operation (i.e.
when the the state $|rr\rangle$ is occupied) there are large
mechanical effects, due to the dipole-dipole force, which create
unwanted entanglement between the internal and the external
degrees of freedom. These problems can be overcome by assuming
single-qubit addressability and by moving to the opposite regime
of small Rabi frequencies $\Omega_1(t) \neq \Omega_2(t)\ll u$. The
gate operation is then performed in three steps, by applying: (i)
a $\pi$-pulse to the first atom, (ii) a $2 \pi$-pulse (in terms of
the unperturbed states) to the second atom, and, finally, (iii) a
$\pi$-pulse to the first atom. The state $|00\rangle$ is not
affected by the laser pulses. If the system is initially in one of
the states $|01\rangle$ or $|10\rangle$ the pulse sequence
(i)-(iii) will cause a sign change in the wave function. If the
system is initially in the state $|11\rangle$ the first pulse will
bring the system to the state $i |r1\rangle$, the second pulse
will be {\em detuned} from the state $|rr \rangle$ by the
interaction strength $u$, and thus accumulate a {\em small} phase
$\tilde \varphi \approx \pi \Omega_2 / 2 u \ll \pi$. The third
pulse returns the system to the state $ e^{i (\pi- \tilde
\varphi)} |11\rangle$, which realizes a phase gate with
$\varphi=\pi-\tilde \varphi \approx \pi$ (up to trivial single
qubit phases). The time needed to perform the gate operation is of
the order $\tau \approx 2\pi/ \Omega_1 + 2\pi/\Omega_2$. Loss from
the excited states $|r\rangle_j$ is small provided $\gamma \Delta
t \ll 1$, i.e.~$\Omega_j\gg \gamma$. A further improvement is
possible by adopting chirped laser pulses with detunings
$\delta_{1,2}(t)\equiv\delta(t)$ and {\em adiabatic} pulses
$\Omega_{1,2}(t)\equiv\Omega(t)$, i.e., with a time variation slow
on the time scale given by $\Omega$ and $\delta$ (but still larger
than the trap oscillation frequency), so that the system
adiabatically follows the dressed states of the Hamiltonian $H_I$.
As found in \cite{jaksch00}, in this adiabatic scheme the gate
phase is \be \varphi(\tau)=\int_{t_0}^{t_0+\tau}\!\!dt\left[{\rm
sgn}(\tilde \delta) \frac{|\tilde \delta| - \sqrt{\tilde
\delta^2+2 \Omega^2}}2-{\rm sgn}(\delta)\left(|\delta| -
\sqrt{\delta^2+\Omega^2}\right)\right]\ee with $\tilde \delta =
\delta -\Omega^2/(4 \delta + 2 u)$ the detuning including a Stark
shift. For a specific choice of pulse duration and shape
$\Omega(t)$ and $\delta(t)$ we achieve $\varphi(\tau)=\pi$. To
satisfy the adiabatic condition, the gate operation time $\tau$
has to be approximately one order of magnitude longer than in the
other scheme discussed above. In the ideal limit $\Omega_j\ll u$,
the dipole-dipole interaction energy shifts the doubly excited
state $|rr\rangle$ away from resonance. In such a
``dipole-blockade'' regime, this state is therefore never
populated during gate operation. Hence, the mechanical effects due
to atom-atom interaction are greatly suppressed. Furthermore, this
version of the gate is only weakly sensitive to the exact distance
between the atoms, since the distance-dependent part of the
entanglement phase $\tilde \varphi \ll \pi$. For the same reason,
possible excitations in the particles' motion do not alter
significantly the gate phase, leading to a very weak temperature
dependence of the fidelity.

In principle the controlled implementation of the above single and
two-qubit quantum gates in the scalable setup of an optical
lattice - avoiding any decoherence - allows any quantum
computation to be performed, and in particular to run Shor's
algorithm \cite{Shor} for factoring large numbers in polynomial
time. We note that the implementation of this factoring algorithm
is one of the major practical goals in quantum computing as most
current cryptographic systems rely on the difficulty of this task.
However, for quantum computers to exceed classical ones in tasks
like factoring they would have to operate thousands of qubits and
the technology for achieving this goal might only be reachable in
years from now. Still, there are also important nontrivial short
term goals which can be realized with nowadays technology for
quantum information processing. We present one of them in the
following section.

\subsection{Feynman's universal quantum
simulator}

A universal quantum simulator (QS) is a controlled device which
efficiently reproduces the dynamics of any other many-particle
system that evolves according to short range interactions.
Therefore a QS could be used to efficiently simulate the dynamics
of a generic many-body system, and in this way function as a
fundamental tool for research in many body physics. Examples
include spin systems, where classical simulations become
nontrivial for a few tens of spins, and also interacting electrons
moving in a magnetic field in 2D, which show interesting effects
like the Fractional Quantum Hall effect. The simulation of such
nontrivial systems is well suited for atoms trapped in an optical
lattice and might therefore be implementable with resources
available in the lab in the near future.

The realization of 2D motion in a 3D optical lattice is
straightforward. Since the height of the barriers between sites
adjacent in $z$-direction increases with the intensity of the
corresponding pair of laser beams one can turn off any motion in
$z$-direction by using high intensities for these lasers while
keeping the barrier between sites adjacent in $x,y$ direction low.
Atoms will then only be able to tunnel along the $x$ and $y$
direction and will thus be confined to one layer of the optical
lattice. The physical simulation of a magnetic field in $z$
direction is less simple since time reversal symmetry needs to be
broken. However, by accelerating the lattice and using additional
laser beams as shown in \cite{Hoflatt} the simulation of a
magnetic field can be achieved and allows to study 2D atom gases
in effective magnetic fields. As was shown by Hofstaedter
\cite{Hof} for strong magnetic fields the energy bands form a
fractal structure depending on the parameter $\alpha = A B e / 2
\pi \hbar$, where $A$ is the area of one of the elementary cells
of the lattice, $B$ is the strength of the magnetic field, and $e$
is the charge of the particles. Plotting the eigenenergies as a
function of $\alpha$ yields the so called Hofstaedter butterfly
shown in \fir{fig11}. While for electrons moving on a 2D surface
it is very difficult to reach sufficiently large values of
$\alpha$ to see this butterfly experimentally the parameter
$\alpha$ can be varied easily between $0$ and $1$ in an optical
lattice and thus it should be possible to observe the energy band
structures shown in \fir{fig11}.

\begin{figure}[tb]
\begin{center}
\includegraphics[width=10.0cm]{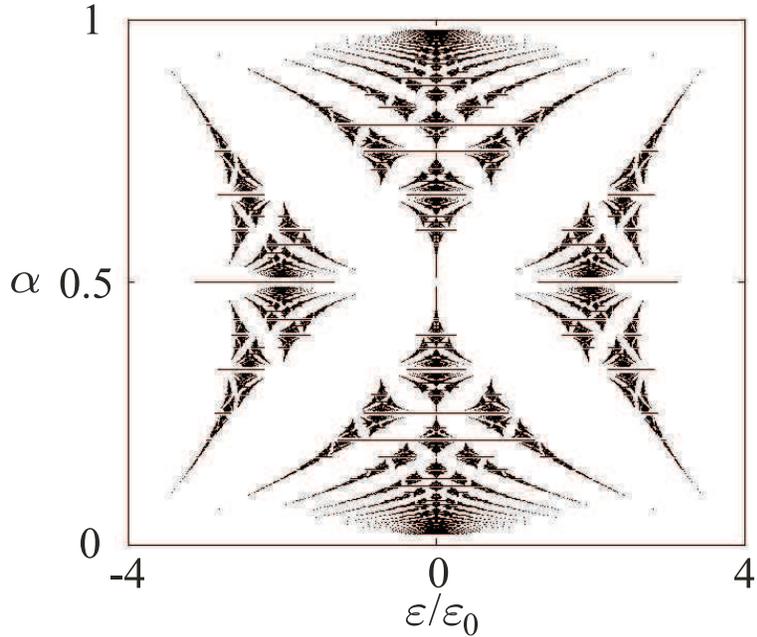}
\caption{The Hofstaedter butterfly. Each dot in this fractal
indicates an eigenenergy of the atoms moving in a magnetic field
with parameter $\alpha$ as given in the text. The energy is
normalized to $\epsilon_0$ which is given by the modulus of the
hopping energy between two sites.} \label{fig11}
\end{center}
\end{figure}

The first idea for simulating (anti)ferromagnetic spin systems
\cite{Molmer} in optical lattices came about after realizing that
the interactions between atoms in optical lattices can be
controlled \cite{jaksch99,Brennen}. In a more general approach it
was recognized by Jane {\em et al.} \cite{Jane} that the nature of
the controlled Hamiltonians available in optical lattices (and
other quantum optical systems) is very well suited for simulating
the time evolution of generic spin chains which are built up by
qubits and evolve according to a Hamiltonian
 \be
H_{N}=\sum_{\ell}H^{(\ell)}+\sum_{\ell \neq \jmath}H^{(\ell
\jmath)}
 \ee
that decomposes into one-qubit terms $H^{(\ell)}$ and two-qubit
terms $H^{(\ell \jmath)}$. It can be shown that the time evolution
resulting from $H_N$ is decomposable into a series of short pieces
$K_n$ which are generated by single- and two-qubit evolutions. The
single qubit evolution is realized in an optical lattice by
properly shining a laser on the trapped atoms and two-qubit
evolutions are obtained by operating two-qubit quantum gates via
atom-atom interactions.

A more detailed investigation shows that interesting spin-chain
Hamiltonians like the (anti)ferromagnetic Heisenberg Hamiltonian
can be simulated without the need for local addressing of single
atoms and could thus be realized with standard optical lattice
setups. The ability to perform independent operations on each of
the qubits would allow all possible bipartite Hamiltonians to be
simulated and in particular the study of quantum phase transitions
in spin chains. We will next briefly study an optical lattice
setup with Rydberg interactions between adjacent atoms which shows
a quantum phase transition and could be viewed as a quantum
simulator \cite{Dorner}. We will also show how it can be used to
implement an atom interferometer whose sensitivity is enhanced by
the created entanglement. Furthermore, the ideas presented in the
next section are the basis for implementing single qubit gates on
robust qubits as introduced in Sec.~\ref{Qmemsinqub}.

\subsection{Multi particle maximally entangled states in optical
lattices} \label{App2}

Let us assume that the Rydberg interactions between neighbouring
atoms are turned on all the time and that the atoms are arranged
in a 1D chain. Furthermore we allow each atom to either tunnel
between two adjacent wells whose ground states are denoted by
$\st{a}$ and $\st{b}$ or to have an additional laser driving Raman
transitions between two internal states (again denoted by $\st{a}$
and $\st{b}$). This situation is schematically shown in
\fir{fig10}.

The dynamics of this setup can be understood as follows. Hopping
between the two modes $\st{a}_j$ and $\st{b}_j$ of the $j$-th atom
is described by
 \be
 H_h=B \sum_j \left( \st{a}_j \bra{b} + {\rm h.c.} \right) \equiv B
 \sum_j \sigma_x^{(j)}
 \ee
with $\sigma_x^{(j)}$ the Pauli $x$-matrix for the two states
(viewed as a spin) of the $j$-th atom and $B>0$ the energy
associated with this process. The ground state of this Hamiltonian
is obtained by putting each atom into an equal superposition of
states written in spin notation as
$\st{\downarrow_x}_j=(\st{\uparrow_z}-\st{\downarrow_z})/\sqrt{2}$,
where $\st{\uparrow_z} \equiv \st{a}$ and $\st{\downarrow_z}
\equiv \st{b}$ i.e.~we find
 \be
 \st{\Psi_h}=\st{\downarrow_x\downarrow_x\downarrow_x \cdots
 \downarrow_x}
 \ee
Note that this part of the Hamiltonian is identical to a spin
chain in a magnetic field $B$ along the $x$-axis.

The interaction between two nearest neighbours due to Rydberg
interactions is given by
 \be
 H_i=W \sum_j \sigma_z^{(j)} \sigma_z^{(j+1)}
 \ee
accounting for the energy difference of having two adjacent
particles in the same vs. different internal states. In this case
the ground state depends on the sign of the interaction parameter
$W$. For repulsive interactions $W>0$ the interaction will be
minimized by arranging the particles as
 \be
 \st{\psi_{ir}}= \alpha \st{\uparrow_z \downarrow_z \cdots \uparrow_z
 \downarrow_z} + \beta \st{\downarrow_z \uparrow_z \cdots
 \downarrow_z \uparrow_z},
 \ee
while for attractive interactions the ground state is
 \be
 \st{\psi_{ia}}= \alpha \st{\uparrow_z \uparrow_z \cdots \uparrow_z}
 + \beta \st{\downarrow_z \downarrow_z \cdots \downarrow_z}.
 \ee
Both of these states are maximally entangled multi particle
states, i.e.~extensions of the well known GHZ states for three
particles.

In the total Hamiltonian $H=H_h+H_i$ the interaction energy and
the hopping energy compete with each other resulting in a quantum
phase transition. When the interaction energy is kept constant and
the hopping term is switched off adiabatically, as e.g.~shown in
\fir{fig10} the state of the system will dynamically change from
$\st{\Psi_h}$ which is a product state to one of the two states
$\st{\Psi_{ir}}$ or $\st{\Psi_{ia}}$ depending on the sign of the
interaction. The exact values of the parameters $\alpha$ and
$\beta$ depend on the details of the dynamics and are discussed in
\cite{Dorner}. These maximally entangled states can serve
different purposes depending on the sign of $W$. Let us discuss
the possible applications of these two kinds of maximally
entangled states

\begin{figure}[tb]
\begin{center}
\includegraphics[width=10.0cm]{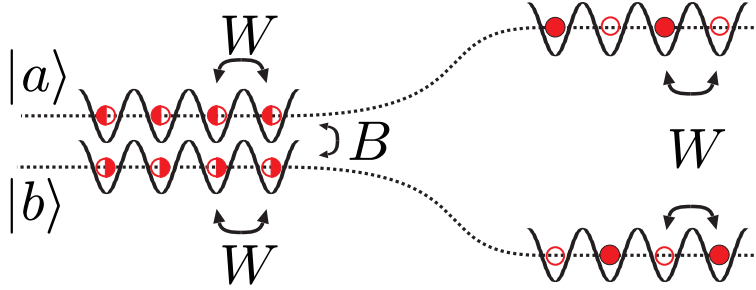}
\caption{Creation of robust multi particle entangled states in 1D
beam splitter setups. Closed circles indicate an atom open circles
an empty position. \label{fig10}}
\end{center}
\end{figure}

\subsection{Repulsive interactions}

In this case both parts of the superposition state have the same
number of particles in each of the two internal states and thus
external stray fields act identically on both, therefore not
affecting the parameters $\alpha$ and $\beta$. Because of this
stability one can use them for storing quantum information in a
robust way as described in Sec.~\ref{Qmemsinqub}. Single qubit
gates can be performed by dynamically going back and forth through
the quantum phase transition in the whole chain changing the
parameters $\alpha$ and $\beta$ in a controlled way.

\subsection{Attractive interactions}

For attractive interactions the terms in the superposition of
$\st{\psi_{ia}}$ will respond to external fields very differently.
Therefore the relative phase between the parameters $\alpha$ and
$\beta$ will be very susceptible to these fields, in fact for $N$
particles in the chain this phase will be $N$ times larger than if
there was just a single particle. The two parts of
$\st{\psi_{ia}}$ can thus be used as two arms of an entanglement
enhanced atomic interferometer.

\section{Summary}

This article reviewed recent advances in controlling and
manipulating neutral atoms by quantum optical methods
concentrating on the use of optical lattices. We investigated
novel physical features arising from interactions in ultracold
dense clouds of atoms as provided by BEC and showed how they
enable experimentalists to gain control over the spatial degrees
of freedom of individual atoms. We also studied possible new
applications - in particular in quantum information processing -
which make use of this unprecedented quantum control. We showed
that universal sets of quantum gates can be implemented and how
they might lead to a universal scalable quantum computer in the
future. Furthermore we also investigated possibilities for shorter
term applications of optical lattices as quantum simulators for
strongly correlated systems.

{Dr. Dieter Jaksch obtained his PhD from the University of
Innsbruck in 1999. He moved to the University of Oxford where he
currently holds a lecturer position at the Clarendon Laboratory in
the Department of Physics in 2003. Dieter Jaksch started to work
on ultracold gases in optical lattices in 1997, his current
interests are mainly in quantum computing applications of quantum
optical systems.}

\end{document}